\documentclass[12pt]{article}
\usepackage{graphicx}

\newcommand\pubnumber{Article 5 in eConf C1304143}
\newcommand\pubdate{\today}

\def\Title#1{\begin{center} {\Large #1 } \end{center}}
\def\Author#1{\begin{center}{ \sc #1} \end{center}}
\def\Address#1{\begin{center}{ \it #1} \end{center}}

\newcommand\pubblock{\rightline{\begin{tabular}{l} \pubnumber\\
         \pubdate  \end{tabular}}}
\newenvironment{Abstract}{\begin{quotation}  }{\end{quotation}}
\newenvironment{Presented}{\begin{quotation} \begin{center}
             PRESENTED AT\end{center}\bigskip
      \begin{center}\begin{large}}{\end{large}\end{center} \end{quotation}}
\def\Acknowledgements{\bigskip  \bigskip \begin{center} \begin{large}
             \bf ACKNOWLEDGEMENTS \end{large}\end{center}}

\def\beq{\begin{equation}}
\def\eeq#1{\label{#1}\end{equation}}
\def\eeqn{\end{equation}}

\def\beqa{\begin{eqnarray}}
\def\eeqa#1{\label{#1}\end{eqnarray}}
\def\eeqan{\end{eqnarray}}

\let\bar=\overbar

\def\Dslash{\not{\hbox{\kern-4pt $D$}}}
\def\dslash{\not{\hbox{\kern-2pt $\del$}}}

\def\msb{{\bar{\ssstyle M \kern -1pt S}}}

\begin{document}
\begin{titlepage}
\pubblock

\vfill \Title{Comparing the observed properties of the GRBs
detected by the Fermi and Swift satellites} \vfill \Author{ L. G.
Bal\'azs$^{1,3}$; I. Horv\'ath$^2$; Z. Bagoly$^{2,3}$  and  J.
K\'obori$^3$ } \Address{$^1$MTA CSFK Konkoly Observatory,
Budapest, Hungary; $^2$National University of Public Service,
Budapest, Hungary; $^3$E\"otv\"os University, Budapest, Hungary }
 \vfill
\begin{Abstract}
We studied the distribution of the GRBs, observed by the Fermi
satellite, in the multidimensional parameter space consisting of
the duration, Fluence, Peak flux and Peak energy (if it was
available). About 10\% of the Fermi bursts was observed also by
the Swift satellite. We did not find significant differences
between the Peak flux and Peak energy of GRBs observed and not
observed also by the Swift satellite. In contrast, those GRBs
detected also by the Swift satellite had significantly greater
Fluence and duration. We did a similar study for the GRBs detected
by the Swift satellite. About 30\% percent of these bursts was
also measured by the Fermi satellite. We found a significant
difference  in the Fluence, Peak flux and Photon index but none in
duration. These differences may be accounted for the different
construction and observing strategy  of the Fermi and Swift
satellites.
\end{Abstract}
\vfill
\begin{Presented}
Huntsville Gamma Ray Burst Symposium, GRB 2013, Nashville,
Tenesse, USA
\end{Presented}
\vfill
\end{titlepage}
\def\thefootnote{\fnsymbol{footnote}}
\setcounter{footnote}{0}

\section{Introduction}
The Swift (launched in 2004) and the Fermi (launched in 2008)
satellites have revolutionized the high energy astronomy in
different manners: Swift is capable to detect GRBs and measure
their  gamma, X-ray and optical properties, along with their
celestial position of high accuracy on the cost of a limited FoV
and  the nominal $\gamma$ spectral range of the BAT instrument is
15-150 keV \cite{BAT}. In contrast, the Fermi satellite has the
GBM telescope covering the whole sky simultaneously, except that
part eclipsed by the Earth, energy response up to 1 MeV with the
NaI detectors (30 MeV with BGO detectors) \cite{GBM} but on the
cost of a limited accuracy of getting spatial position of the
GRBs. A very attractive feature of the satellite is the LAT
\cite{LAT} capable to localize and detect photons above the energy
of 10 MeV. Although, the energy range of the Swift and Fermi GBM
sensitivity has a significant overlap, the differences in
detection strategy and construction may result deviations between
the GRB populations observed by these two experiments. Therefore,
it is an interesting issue to make comparison between GRBs
detected by both satellites and those registered only one of these
two instruments. Due to the different geometry of the orbits of
the two satellites it could well happen that some of the bursts
were in the FoV at only one of them. In the following we shortly
address this issue. We used the data in the Fermi GBM Catalogue
\cite{fermi} and Swift GRB  Table \cite{swift}.

\section{Coincidence between Fermi and Swift detections}
\begin{figure}[htb]
\centering
\includegraphics[height=4in]{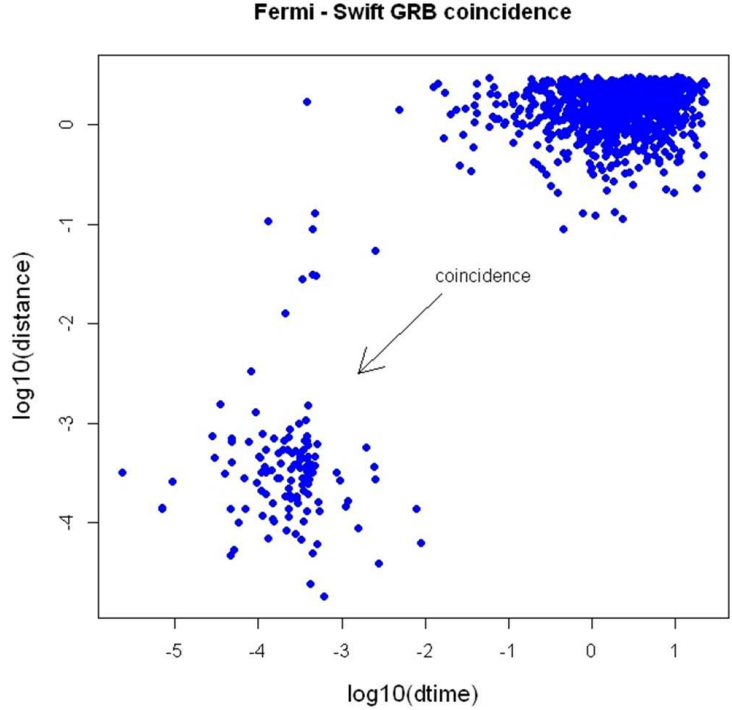}
\caption{Coincidence  between  the  Fermi and Swift detections.
The point cluster  at the left lower side of the  scatter plot is
considered  as bursts jointly detected by the Fermi and Swift
satellite } \label{fig:FSW}
\end{figure}

From the very beginning of the Fermi experiment in 2008 the GBM
detected 1070 bursts until mid of February, 2013, starting the
necessary statistical computations. In the same period Swift
recorded 409 GRB events. Taking the Fermi bursts we searched for
the nearest Swift burst in time and position and computed the
differences. Unfortunately, for the majority of bursts detected by
the GBM the accuracy of the celestial position was in the order of
a degree which is too coarse for getting a reliable spatial
coincidence. We display the scatter plot between the time and
spatial difference to the nearest Swift burst in Figure
\ref{fig:FSW}. It is obvious already at the first glance of this
Figure that the points forms two completely separated clusters. It
is also evident from this Figure that points seem to coincide with
Swift detection time may deviate considerably in the Fermi
position. A good example for this is the group of points between
the two major concentrations. We considered  those bursts to be
detected by both satellites which fulfilled the $dtime$ $<
10^{-3}$ $day$ and $distance$ $< 10^{-3}$ $radian$ conditions. In
this way out of the 1070 burst detected by Fermi only 115 could be
considered as also detected by the Swift satellite.

\section{Fermi GRBs detected/non detected by the Swift}

Since the two satellites has different spectral responses and
sensitivities it is an appropriate question whether is there some
systematic difference between bursts detected only by the Fermi or
jointly by the Swift satellite. We compared the duration, fluence,
peak flux and  spectral peak energy by means of Kolmogorov-
Smirnov (KS) tests. In  Table \ref{FSKS} bold face means
significant differences.

\begin{table}[t]
\begin{center}
\begin{tabular}{r|cccc}
\   & Fluence & t90 & P1024 & Epeak  \\
\hline
Most Extr.  Diff.  Absolute & .194  & .145 & .107  & .096  \\
                 Positive & .194 &  .145 &  .107 &  .051 \\
                Negative & -.003 & -.022 & -.006 & -.096  \\
      Kolmogorov-Smirnov Z & 2.018 & 1.511 & 1.111  & .688  \\
\hline
         Asymp. Sig. (2-tailed) & {\bf .001} & {\bf .021} & .169 & .730  \\
\hline
\end{tabular}
\caption{KS test statistics of Fermi GRBs detected/non detected by
the Swift. Bold face means significant difference.}
\label{FSKS}
\end{center}
\end{table}

Table shows that no difference can be detected in the Peak
intensity and the spectral Peak energy. The Peak intensity is the
primary physical quantity responsible for the detection.  If it
does not exceed the noise level significantly, the burst is not
detected. Since there is no significant difference between the
jointly and only by the Fermi detected bursts in the Peak flux,
the low fraction of the joint detections may be explained by the
much greater FoV of Fermi. The significant differences in t90 and
Fluence may originate from higher sensitivity of Fermi for GRBs of
short duration.

\section{Swift GRBs detected/non detected by Fermi}
In the same period in which we considered the Fermi GRBs Swift
recorded 409 bursts. The Swift is working in observatory mode,
consequently, its FoV is significantly smaller. On the contrary,
it has higher sensitivity. Due to its construction Swift BAT has
smaller beam size than the Fermi GBM. Assuming the same background
level at both satellites the smaller beam size results in smaller
noise contribution to the burst event to be observed. For making a
comparison between GRBs detected jointly by the BAT and GBM  and
by the BAT alone we selected  Fluence, t90, Peak flux and Photon
index. We compared these variables between the non detected and
jointly detected groups by performing KS tests. Table \ref{SFKS}
summarizes the results.

\begin{table}[t]
\begin{center}
\begin{tabular}{r|cccc}
    & t90   & Fluence & Peak    & Pind \\
\hline
Most  Extr. Diff.     Absolute &    .085    & .279  & .332  & .170 \\
       Positive & .078  & .279  & .332  & .022 \\
      Negative  & -.085 & -.013 & -.005 & -.170 \\
       Kolmogorov-Smirnov Z &   .752    & 2.521 & 2.976 & 1.540 \\
\hline
          Asymp. Sig. (2-tailed) &  .624    & {\bf .001}$>$ & {\bf .001}$>$ & {\bf .017} \\
\hline
\end{tabular}
 \caption{KS test statistics of Swift GRBs detected/non
detected by Fermi. Bold face means significant difference.}
\label{SFKS}
\end{center}
\end{table}

\begin{figure}[htb]
\centering
\includegraphics[height=4in]{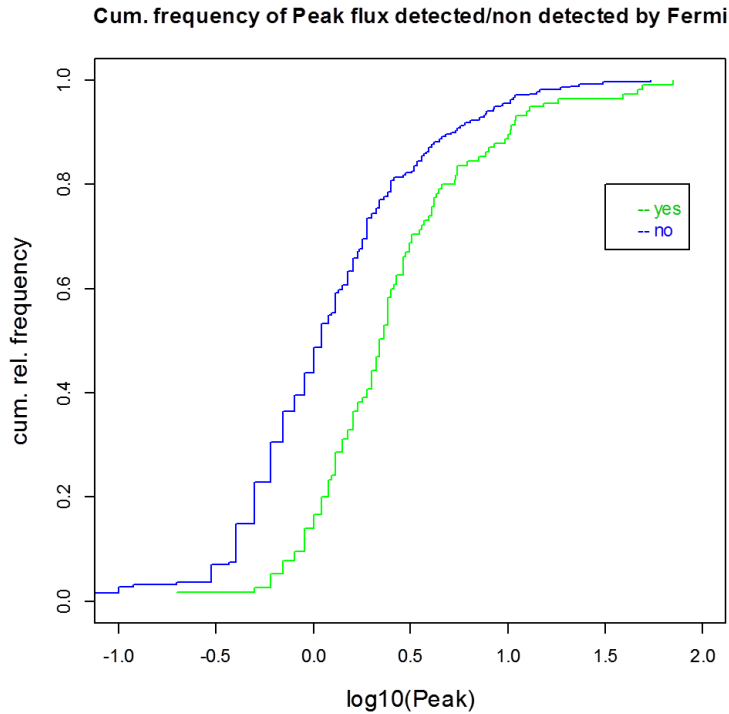}
\caption{Comparison of the cumulative distribution of Swift GRB's
peak fluxes detected/non-detected (green/blue) by the Fermi
satellite. KS test indicates a significant difference at the
$>99.9\%$ level.} \label{fig:SFW}
\end{figure}

The Table shows that except the t90 duration there are significant
differences between the Fluence, Peak flux and Photon index. The
strongest difference appears in the Peak flux. The reason for this
difference may be accounted for the higher sensitivity of the BAT
instrument. To demonstrate this difference we displayed the
cumulative distribution of the Peak flux in the non detected and
jointly detected group in Fig. \ref{fig:SFW}.

Figure \ref{fig:SFW} demonstrates convincingly that about 30\% of
the GRBs, detected by the BAT only, have fainter Peak flux than
those recorded also by the GBM.  On the other hand, it means that
70\% of the BAT detected GRBs are above the threshold of GBM. An
obvious reason of this contradiction is the different orbital
position of the two satellites, and consequently the different
part of the sky covered by the Earth (unfortunately, the detailed
study of this effect is already beyond the scope of this work).

\section{Summary and Conclusions}
We compared the GRB detections of the Swift and the Fermi
satellites. We studied the period from the  beginning of the Fermi
mission until middle of February 2013. In this period Fermi
registered 1070 burst while Swift recorded 409 events. We
considered the GRBs as jointly observed by both satellites if the
time and positional difference was less than $10^{-3}$ day and
$10^{-3}$ radian, respectively. We obtained 115 bursts fulfilling
these criteria. Using Kolmogorov-Smirnov (KS) tests we compared
the distributions of the durations, Fluence, Peak flux and Epeak
of the bursts which were jointly detected by both satellites and
by the Fermi alone. We did not get significant difference between
the two category of the bursts in the Peak flux and Epeak
distributions. On the other hand, a significant difference can be
obtained in the Fluence and duration. This effect may be accounted
for the higher sensitivity of Fermi to the bursts of short
duration (t90 $<$ 2s).

A similar comparison of the duration, Fluence, Peak flux and
Photon
 index of bursts of jointly detected and by the Swift alone resulted
 in significant differences utilizing KS tests, except the duration.
 The most significant difference was obtained in the Peak flux.
 Comparing the cumulative distribution of the Peak flux of the two
 categories (jointly detected and by the Swift alone) demonstrated
 that about 30\% of the Swift GRBs is below the detection limit of
 Fermi. Since 70\% of the bursts detected by the Swift alone is in
 the sensitivity range of Fermi the non detection may caused by
 the differences in the orbital position at observing a given burst.

Summarizing all these things we concluded that the differences in
detections of the Fermi and Swift satellites may be accounted for
the different construction and detection strategy, along with the
different orbital positions at a particular burst event.

\Acknowledgements This research was supported by  OTKA grant
K77795, by OTKA/NKTH A08-77719 and A08-77815 grants (Z.B.).

\end{document}